# Ageing dynamics of colloidal hard sphere glasses

V.A. Martinez#, G. Bryant and W. van Megen

*Department of Applied Physics, Royal Melbourne Institute of Technology, Melbourne,*

*Victoria 3000, Australia*

ABSTRACT

We report results of dynamic light scattering measurements of the coherent intermediate scattering function (ISF) of glasses of colloidal hard spheres for several volume fractions and a range of scattering vectors around the primary peak of the static structure factor. The ISF shows a clear crossover from an initial fast decay to a slower non-stationary decay. Ageing is quantified in several different ways. However, regardless of the method chosen, the perfect "aged" glass is approached in a power law fashion. In particular the coupling between the fast and slow decays, as measured by the slope at the crossover, also decreases algebraically with waiting time. The non-stationarity of this coupling implies that even the fastest detectable processes are themselves non-stationary.

# Present address; School of Physics and Astronomy, The University of Edinburgh, Mayfield Road, Edinburgh, EH9 3JZ, United Kingdom

# 1. INTRODUCTION

Ageing—the non-stationary dynamics of deeply quenched fluids—has been a subject of considerable investigation over the last decade. Examples of materials that have been studied recently include sponges, foams and polymer and colloidal gels [1]. Unlike window glass and obsidian, these mesoscopic, soft complex materials lend themselves particularly well to the study of aging; their non-stationary structural dynamics are commonly exposed by the dependence of the time correlation function of density fluctuations, (the intermediate scattering function (ISF)), on the waiting time. Despite the diversity and complexity illustrated by these examples, their dynamics share several features; An initial fast decay of the ISF to a plateau is followed by a slower decay. The initial decay is observed to be independent of age, while the decay from the plateau is observed to slow as the material ages. The fact that this slow decays follows a compressed exponential function of the delay time suggests an underlying very slow ballistic, rather than diffusive, motion. And indeed, these features have been attributed to elastic deformation driven by the evolution or redistribution of stresses, such as occurs in the shrinkage of gels and coarsening of foams [2].

Stationarity of the fast process and scaling of the slow, non-stationary decay with waiting time are two aspects that these experimental observations share with the results of numerous computer simulations of simple molecular glasses [3,4]. The difference is that the ISFs of the molecular glasses tend to decay algebraically from their respective plateaux. Quenched-in stresses appear to be relieved through cooperative, avalanche-like events [3].

A suspension of particles with hard sphere-like interactions presents a simpler experimental system. The interactions between the particles, deriving from thin solvated oligomeric surface layers, are effectively ageless. Provided the spread of particle radii is not too large—less than about 10% of their mean—these suspensions show a transition from a disordered, fluid-like phase to an opalescent, crystal phase that mimics the freezing-melting transition of simple atomic materials [5]. They also show a glass transition (GT), identified by the particle concentration where density fluctuations fail to fully relax on any reasonable experimental time scale [6]. Even in the context of the GT the existence of the first order transition, in that it provides a thermodynamic reference for locating the under-cooled, non-equilibrium states, remains pivotal.

When the glass of colloidal hard spheres was first observed its ageing was not exposed due, in no small part, to the limited understanding of how to deal experimentally with non-ergodicity and non-stationarity. The light scattering procedures developed at the time to account for non-ergodicity assumed stationarity [6,7]. In that sense the colloidal glass was considered ideal.

Indeed, the transition from colloidal fluid to the glass still appears to be the best experimental illustration of the sharp transition predicted by the idealized version of mode-coupling theory (MCT) [8]— a view reinforced by quantitative consistency between MCT and results of Dynamic light scattering (DLS) experiments for these colloidal systems [9,10]. Neglect of those irreversible, ergodicity restoring processes, processes so much more vigorous in molecular glass formers than in colloidal systems, is generally considered to be the theory's main shortcoming [11]. The incorporation of such processes necessarily requires the introduction of further approximations and assumptions. However, by comparison with experiment the idealized theory can also be exploited to aid in quantifying the very processes it neglects [10,12].

The first technique by which a statistically viable ensemble of density fluctuations, having decay times longer than the measurement time, could be accumulated with sufficient speed to resolve slow non-stationary processes was introduced by Müller and Palberg in 1996 [13]. When this technique was subsequently applied to the glass of colloidal hard spheres its ageing was exposed [14]. However, at the time the process was neither quantified nor analysed in detail.

A preliminary quantitative account of ageing of the glass of hard sphere colloids was presented in a recent Letter [15]. This paper presents not only a more extensive examination of the processes but also details, in the following Section, the protocols used to acquire good estimates of ensemble averages for processes whose decay times exceed any practical measurement time, as well as robust methods for distinguishing between reversible and irreversible (ageing) processes, however slow or fast each of these may be. In Sec. 3 results are presented and discussed for several volume fractions and a range of wavevectors around the main peak of the static structure factor. Conclusions are presented in Sec. 4.

## 2. METHODS

### 2.1 Sample Description

The particles used in these experiments comprise cores of a co-polymer of methylmethacrylate and trifluoroethyl acrylate with coatings of poly(12-hydroxystearic acid) approximately 10nm thick [16]. Average hydrodynamic radii, R, and polydispersities, $\sigma$, listed in Table 1, were determined by DLS and Static Light Scattering (SLS) on very dilute samples [17]. Suspension of the particles in *cis*-decalin achieves matching of the refractive indices of particulate and suspending phases to a degree that effectively suppresses multiple scattering over the range of wave vectors (between $qR \approx 1.5$ and $qR \approx 4$) spanned in these experiments. The reduction in van der Waals attraction this matching achieves, along with the lack of detectable charge on the particles, means that the interactions between them are dominated by the short range repulsion effected by the solvated surface coatings [18]. Hence these suspensions present a good approximation to the perfect hard sphere system. Accordingly, after mapping the equilibrium phase behaviour observed for these suspensions onto that known for the perfect, one-component hard sphere system, the volume fractions $\phi_f \approx 0.493$ and $\phi_m \approx 0.54$ of coexisting fluid and crystal phases can be identified [5]. With this reference the observed glass transition is located at $\phi_g \approx 0.57$ [6].

Alternatively, using the results of recent computer simulations one can employ, in the above manner, the freezing volume fraction, $\phi_f(\sigma)$, of a system of hard spheres with a finite spread, $\sigma$, in the particle size distribution (PSD); From Ref. 19 one reads for example $\phi_f(8\%) \approx 0.527$. It has been argued that comparison between different systems requires the volume fraction to be referenced in this way [20]. However, the available computational data for $\phi_f(\sigma)$ is limited to symmetrical PSDs while those of the type of polymer particles used here tend to be negatively skewed [17]. So, rather than introducing ambiguity due to the influence of the shape of the PSD by attempting to express the volume fraction in absolute terms, and to remain consistent with the vast majority of literature, the stated sample volume fractions will be read here relative to the freezing value, $\phi_f = 0.493$. For comparison, however, volume fractions relative to the freezing point for a symmetrical PSD are provided in Table 1. Alternatively, the thermodynamic phase point may be expressed in terms of the degree of under-cooling, or over-packing in this case, $\varepsilon = (\phi/\phi_f - 1)$. Two different latex suspensions are used in this work. Samples designated XL52 are used for the validation of the techniques and the q dependence, and samples designated XL63 for the remainder of the paper (the volume fractions, polydispersities and scattering vectors studied, are given in table 1). Despite the slight differences between the two latices the DLS results are consistent.

| Suspension | $\phi$ | $\phi_p$ | $\varepsilon$ | qR |
|---|---|---|---|---|
| **XL52, R = 200 nm** <br> **$\sigma = 9\%$, $\tau_b = 0.0193$ s** | 0.60 | 0.637 | 0.217 | 2.01, 2.93, 3.57, 4.15 |
| **XL63, R = 185 nm** <br> **$\sigma = 8\%$, $\tau_b = 0.0153$ s** | 0.563 | 0.595 | 0.142 | 1.86, 2.71, 3.30, 3.84 |
| | 0.574 | 0.606 | 0.164 | 3.30 |
| | 0.584 | 0.616 | 0.185 | 2.71, 3.30, 3.84 |
| | 0.600 | 0.631 | 0.217 | 1.86, 2.71, 3.30, 3.84 |

Table 1. Volume fractions and scattering vectors for the two types of particle used in these experiments. First column; volume fractions, $\phi$, referenced to the freezing value of the one component hard sphere system and used throughout this paper. Second column, volume fractions, $\phi_p$, referenced to the freezing value of the polydisperse hard sphere system. See text for further details.

Prior to the start of a measurement each sample was tumbled for at least 24 hours, effectively shear melting any colloidal crystals that may be present. The waiting time origin, $t_w = 0$, is the time a sample was placed in the spectrometer immediately after tumbling was stopped. In the results below delay times, $\tau$, are expressed in units of the Brownian time, $\tau_B = R^2/(6D_o)$, where $D_o$ is the diffusion constant for freely diffusing particles [Table 1]. Waiting times, $t_w$, are expressed in hours.

## 2.2 Light Scattering Procedures

The quantity of interest, and that from which other dynamical properties are derived in this paper, is the intermediate scattering function or auto-correlation function of the $q^{th}$ spatial Fourier component, $\rho(q,\tau)$, of the particle number density fluctuations,

$$f(q,\tau) = \left\langle \rho(q,0)\rho^{*}(q,\tau) \right\rangle \Big/ \left\langle \left| \rho(q) \right|^{2} \right\rangle, \tag{1}$$

where the asterisk denotes the complex conjugation.

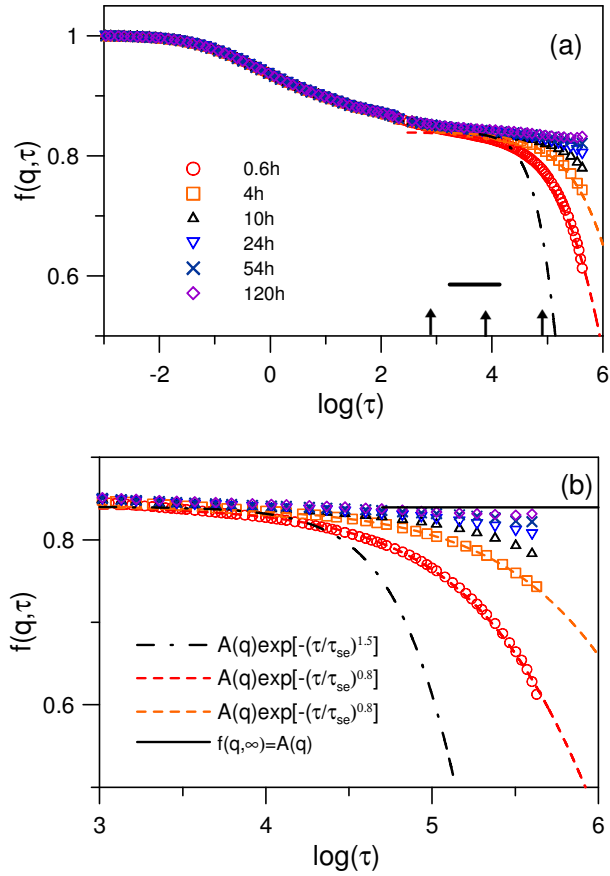

Fig. 1. ISF versus logarithm of delay time, for qR=2.93 and $\phi$=0.60, at waiting times indicated. Arrows in (a) indicate measurement times, $T_l/\tau_b$, and the small horizontal bar indicates range of variation in crossover time, $\tau_m(q)$, with waiting time (see text for details). (b) Expanded scale of delay time highlights quality of stretched exponentials fits; values of $\tau_{se}$ are (left to right) $2.2\times10^{5}$, $2.1\times10^{6}$, $6.5\times10^{6}$.

For the purpose of justifying the protocols adopted we show, in Fig. 1, a typical result for the ISF, $f(q,\tau)$, for several waiting times, $t_w$. One sees first, that $f(q,\tau)$ fails to decay to zero in the experimental time window and second, that $f(q,\tau)$ varies with waiting time, $t_w$. Respectively these indicate non-ergodicity and non-stationarity—features of the glass and complications for the experimentalist not normally associated with the fluid state. One also notes that a plateau separates an initial (fast) decay from a slow decay that stretches out as $t_w$ increases. Whether the fast process is independent of $t_w$, as it appears to be, is one of the aspects of the relaxation dynamics examined below.

The stretching function

$$n(q,\tau) = \frac{d\log(w(q,\tau))}{d\log\tau}, \qquad (2)$$

where

$$w(q,\tau) = -q^2 \ln f(q,\tau), \qquad (3)$$

is an alternative presentation of the data [12]. One sees from examples shown in Fig. 2 that their minima, $\nu(q)=\min[n(q,\tau)]$, deepen and shift to longer delay times, $\tau_m(q)$, as the sample ages. We identify $\tau_m(q)$ as the crossover time from the fast process to the slow process. It is also the delay time where the non-Fickian, or collective, dynamics are most strongly exposed. Accordingly we give considerable currency to the quantities $\tau_m(q)$, $\nu(\tau_m(q))$ and $f(q,\tau_m(q))$.

One way to deal with non-ergodicity is with the approach introduced by Pusey and van Megen (PvM) [7,21]. This assumes that the space-time density fluctuations can be decomposed into an arrested, time-invariant component and a time dependent, fluctuating component. The resulting scattered light field comprises constant and fluctuating components. Stationarity is implicit. Then, provided the spatial correlation of both components is much smaller than the linear dimension of the scattering volume, thereby allowing the Central Limit Theorem to be effected, the fluctuating field is a complex, Gaussian variable of zero mean. These considerations lead to the following expression for the ISF [7];

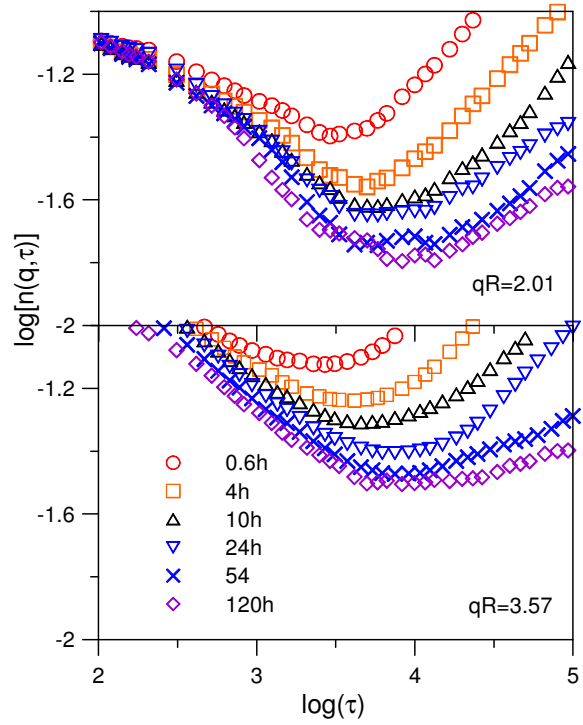

Fig. 2. Double logarithm plots of the stretching functions [Eq. (2) and (3)] for $\phi=0.60$, versus delay time for wavevectors and waiting times indicated.

$$f_1(q,\tau) = 1 + \frac{\langle I \rangle_{T_1}}{\langle I \rangle_E} \left\{ \left[ \frac{\langle I(q,0)I(q,\tau) \rangle_{T_1} - \langle I^2(q) \rangle_{T_1}}{\langle I(q) \rangle_{T_1}^2} + 1 \right]^{1/2} - 1 \right\} . \quad (4)$$

The brackets, $\langle ...... \rangle_{T_1}$, express the time average over the intensity fluctuations, $I(q,t)$, of a single speckle (or spatial Fourier component of the particle number density fluctuations) accumulated in a single measurement of duration $T_1$, and $\langle ...... \rangle_E$ express the ensemble average, acquired by averaging over a large number ($\approx4000$) of speckles, achieved here in a single rotation of the sample. The non-ergodicity factor, $f_1(q,\infty)$ follows from Eq. (4) in the limit $\tau\rightarrow\infty$:

$$f_1(q, \infty) = f_1(q, \tau \to \infty) = 1 + \frac{\langle I(q) \rangle_{T_1}}{\langle I(q) \rangle_E} \left\{ \left[ 2 - \frac{\langle I^2(q) \rangle_{T_1}}{\langle I(q) \rangle_{T_1}^2} \right]^{1/2} - 1 \right\} \qquad (5)$$

The derivation of Eq. (4) assumes a detector area smaller than one coherence area; ie, only a single spatial Fourier component of the scattered light is detected. The subscript "1" is used to denote results obtained by the above procedure.

Another approach is to accumulate the time averages of the intensities, $\langle I(q) \rangle_{T_2}^{(i)}$, and intensity auto correlation functions (ACFs), $\langle I(q,0)I(q,t) \rangle_{T_2}^{(i)}$ for a (large) number, M, of independent spatial Fourier components of the particle number density fluctuations. From these one estimates the (normalized) ensemble-averaged intensity ACF as follows,

$$g_E(q, \tau) = \frac{\frac{1}{M} \sum_i \langle I(q,0)I(q,t) \rangle_{T_2}^{(i)}}{\left[ \frac{1}{M} \sum_i \langle I(q) \rangle_{T_2}^{(i)} \right]^2} \qquad (6)$$

In some of the earliest studies of colloidal glasses [6,9] the "brute force" accumulation implied by Eq. (6) was achieved, rather inefficiently, by either translation or rotation of the sample between individual measurements. Ageing was neither observed nor considered. Accumulation of independent spatial Fourier components of the fluctuations efficient enough to resolve non-stationary processes can now be achieved by either collecting images of the far field intensity distribution (or speckle pattern) on a CCD camera [22], or by gating the intensity of the scattered light collected by a standard PMT while the sample is rotated continuously [13,23]. Either procedure provides an average over some 4000 speckles. Technicalities place lower limits on the respective delay time windows of order 0.1s and 1 s, and this limitation necessitates complementary measurement of faster processes using Eq. (4).

Of the two procedures based on continuous rotation, interleaved sampling [13] and the method of echoes [23], the second is employed here. It has the higher resolution and is able to account for errors in the period of rotation. Presuming the 4000 or so speckles accumulated in a single rotation provides the basis for an adequate estimate of the ensemble averaged intensity ACF, Eq. (6), the ISF can be calculated by the usual Siegert relationship,

$$g_E(q, \tau; t_w) = 1 + c |f_2(q, \tau; t_w)|^2. \qquad (7)$$

Explicit dependence on the waiting time, $t_w$, is denoted here in the expectation that the slower processes, at least, may be non-stationary.

The unknown experimental factor c, in Eq. (7), is fixed by requiring that the results of the two measurements match, ie, $|f_1(q, \tau)|^2 = |f_2(q, \tau; t_w)|^2$, for that range of delay times (around 1s) where the results of the two procedures overlap. So normalization of $f_2(q, \tau; t_w)$ depends on $f_1(q, \tau)$. The subscript, "1" or "2", is then dropped from the final ISF so obtained.

The duration, $T_1$, of measurement of the ISF by the PvM method, was selected so as to maximize inclusion of the fast, and predominantly stationary, processes and minimize the inclusion of slow, non-stationary processes: Thus, the upper limit of the time window of this part of the measurement must be at least comparable to $\tau_m(q)$. The duration, $T_2$ (2000 to 10,000s), of individual measurements by the echo method was selected to best resolve and expose non-stationary processes.

Fig. 3 shows $f_1(q, \tau)$ for values of $T_1$ of 100s, 1000s, and 10000s. The upper limits of the time windows, $T_1/\tau_b$, corresponding to these values of $T_1$ are indicated in Fig. 1. Fig. 4 shows $f_1(q, \tau)$ for values of $t_w$, from 1h to 115h. These two figures indicate that any effects that the measurement time and waiting time may have on the ISF, measured by the PvM approach, are within experimental noise. However, a subtlety is exposed in Fig. 5, where the non-ergodicity factors, $f_1(q, \infty)$, obtained by Eq. (5), are shown as a function of $t_w$ for the above values of $T_1$. As is evident from Eq. (5), the greater the spread, $\langle I^2(q) \rangle_{T_1}$, of intensity fluctuations accumulated in the given measurement, irrespective of whether these derive from stationary or non-stationary processes, the smaller the corresponding value of $f_1(q,\infty)$. So, one sees that increasing $T_1$ from 100s to 10,000s results in a decrease in $f_1(q,\infty)$, averaged over waiting times, of some 3%.

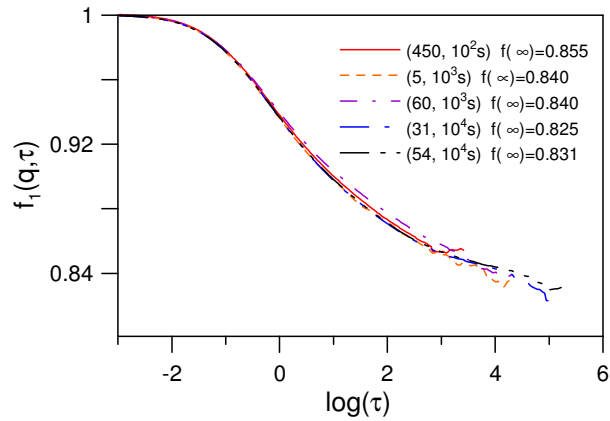

Fig. 3. ISF from PvM method versus logarithm of delay time at qR=2.93 for $\phi$=0.60 for several values of the number of measurements (n) and duration ($T_1$), as indicated in parentheses (n, $T_1$). Note the expanded vertical scale.

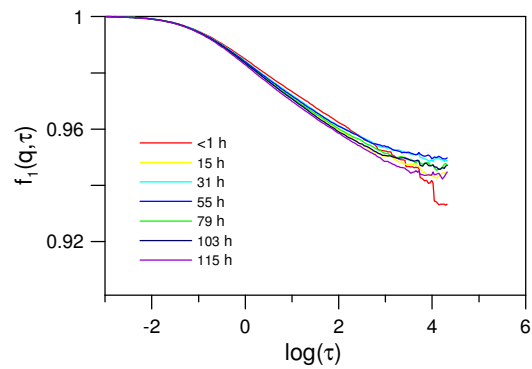

Fig. 4. ISF from PvM method versus logarithm of delay time for several waiting times (qR=3.57 and $\phi$=0.60). Each curve presents an average over 10 measurements of duration 1000s. Note that the value of qR and the scale of the ordinate differ from Fig. 3. Note the expanded vertical scale.

This small variation with $T_1$ is effectively cancelled when the results, obtained by the PvM and echo methods, are averaged in the time window where they overlap. The final ISF, $f(q,\tau;t_w)$, is insensitive to variation of $T_1$ from 100s to 10,000s. In particular $f(q, \tau_m(q))$, the ISF at the crossover is independent of $T_1$ and, as may be seen in Fig.

5, shows no variation with $t_w$. Accordingly, the measurement time $T_1$ (=1000s in this case) selected is that for which $<f_1(q,\infty)>=f(q,\tau_m(q))$; ie the conditions for which the non-ergodicity factor coincides with the ISF at the crossover. This criterion removes the ambiguity, albeit a small one, from our experimental definition of the non-ergodicity factor. The latter is henceforth denoted without subscript as $f(q,\infty)$. Of course for a given set of conditions preliminary measurements must be made in order to estimate $\tau_m(q)$ and $f(q,\tau_m(q))$ approximately before determining the final protocol.

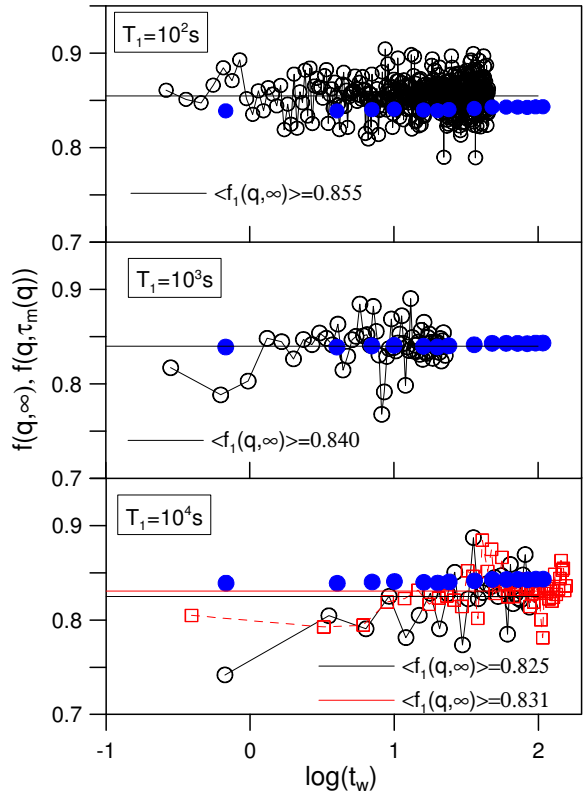

Fig. 5. Non-ergodicity factor, $f_1(q,\infty)$, [Eq.(5)] versus logarithm of the waiting time (hours), shown by open circles and squares, for measurement times, $T_1$, indicated (qR=2.93 and $\phi$=0.60). Open circles and squares seen in the bottom panel derive from independent sets of measurements. Values of these non-ergodicity factors averaged over all waiting times are given and shown by horizontal lines. The ISF, $f(q,\tau_m(q))$, at the crossover, are shown by the full circles .

Finally, we demonstrate that the rotation of the sample does not influence the dynamics. Fig. 6 illustrates that f(q,τ), shown for two delay times, is not affected by either the rotation period or the time, $T_2$, for which the sample is rotated and the measurement performed. Fig. 7a and 7b show that the parameters ν(q) and $τ_m(q)$ are also unaffected by the measurement time or rotation period. It is appropriate here to mention that a recent study [24] of ageing of a water-oil emulsion shows consistency of results obtained by imaging the speckle pattern on a CCD camera (ie, without rotation of the sample) and the method of echoes (with rotation).

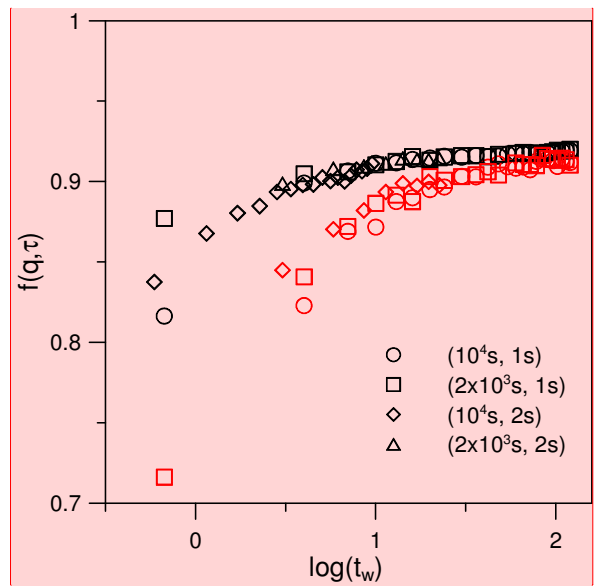

Comment [VM1]: See the table included in the excel file.

Fig. 6. ISF, f(q,τ) versus logarithm of the waiting time at delay times τ=8×10⁴ (upper data set) and τ=4×10⁵ (lower data set). Measurement time, $T_2$, and rotation period, $t_r$, are indicated in parentheses $(T_2, t_r)$.

# 3. RESULTS AND DISCUSSION

## 3.1 Aging of a colloidal glass

The main result from the previous section is that the non-ergodicity factor, f(q,∞), is a stationary property of the colloidal glass investigated here, at least for the wavevectors (listed in Table 1) around the position, given by qR≈3.6, of the main peak of the static structure factor. Values of f(q,∞) obtained in this work are shown in Fig. 8. They are consistent with those of previous studies [25] and MCT [8].

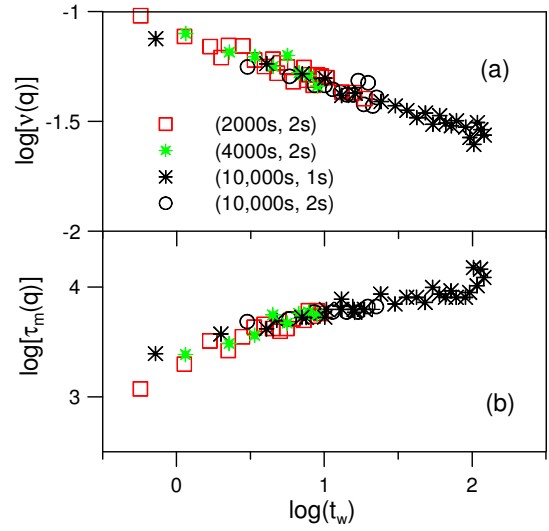

Fig.7. Double logarithm plots of (a) stretching index, ν(q), and (b) cross-over time, $\tau_m(q)$ versus waiting time. Measurement time, $T_2$, and rotation period, $t_r$, are indicated in parentheses ($T_2$, $t_r$).

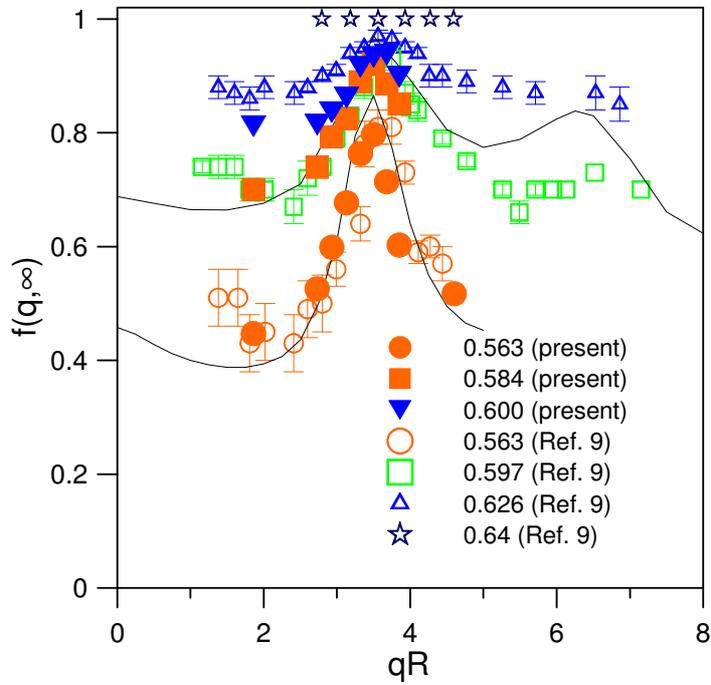

Fig. 8. Non-ergodicity factors, f(q,∞) versus wavevector for volume fractions indicated. Solid symbols are the present data and the open symbols are results from Ref. 25. The solid lines are from MCT [9].

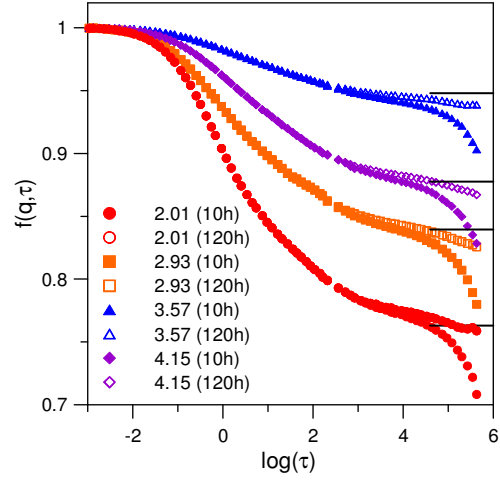

Fig. 9. The ISF versus logarithm of delay time for φ=0.600. for wavectors and waiting times indicated in parentheses. For each case the non-ergodicity parameter, f(q,∞), is indicated by a horizontal line.

Fig. 9 shows the ISFs for several wavevectors. As in Fig. 1, ageing is most pronounced at the upper limit, $\tau_{max}$=4.5×10$^5$, of the time window but almost imperceptible at the crossover. We quantify the aging process by plotting, as in Fig. 10, the ISF as a function of the waiting time, for fixed values of the delay time. The variation with $t_w$ can be most simply described by the power law,

$$f\left(q,\tau;t_w\right) = A(q)\left[1-\left(\frac{t_w}{t_0}\right)^{-b}\right].$$
(8)

Fitting this to the ISFs for the range of delay times from the crossover time, $\tau_m$ (~10$^4$), to $\tau_{max}$ consistently gives the same values for A(q), as seen in Fig. 11a. From the latter one also sees, within experimental uncertainty, that A(q) coincides with the non-ergodicity factor, f(q,∞), and therefore with the ISF, f(q, $\tau_m$(q)), at the crossover. Thus, following the tumbling, an action one might consider as synonymous with a quench, the perfect "aged" glass, that for which the ISF decays to a constant value, say $f(q,\tau>\tau_m(q))=f(q,\infty)=A(q)$, is approached algebraically with waiting time. The other two parameters, b and $t_o$, in Eq. (8) are shown in Fig. 11b and 11c. As indicated, uncertainties are appreciable and any dependence these quantities may have on the wavevector cannot be discerned with confidence. These results are, however, not inconsistent with the expectation that as the delay time is increased a larger amplitude, $t_o^b$, of the non-stationary processes is obtained.

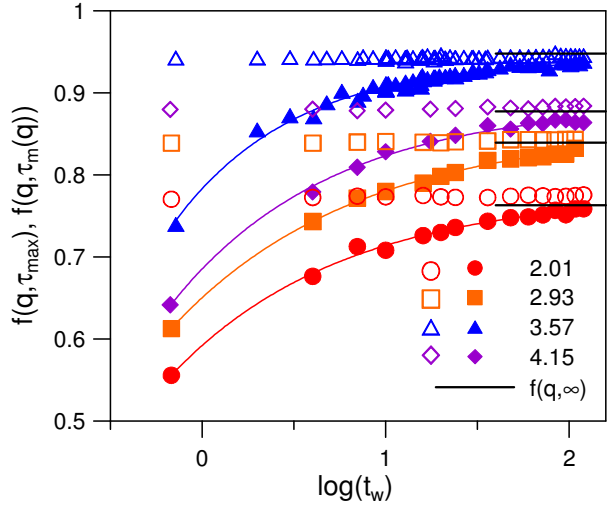

Fig. 10. $f\left(q, \tau_{max} \approx 4.5 \times 10^5\right)$ (filled symbols) or $f\left(q, \tau = 10^4 \approx \tau_m(q)\right)$ (open symbols) as functions of the waiting time for several values of qR indicated. Lines are fits of Eq. (8) to the data. Values of $f(q,\infty)$ are indicated by small horizontal lines.

From Fig. 1 and the discussion in Sec. 2.2 it would appear that the ISF can be neatly and conveniently divided into a fast, stationary component and a slow, non-stationary component. We investigate the cross-over between these components more-closely, first, with the stretching index, which is identified as the minimum in the stretching function (Eq. 2): $\nu(q)=\min[n(q,\tau)]$. Inspection of the waiting time dependence of $\nu(q)$, shown in Fig. 12, exposes a coupling, however weak, between the fast and slow processes. Moreover, as illustrated, this coupling decreases algebraically with waiting time;

$$\nu(q) = \nu_o t_w^{-\zeta}. \tag{9}$$

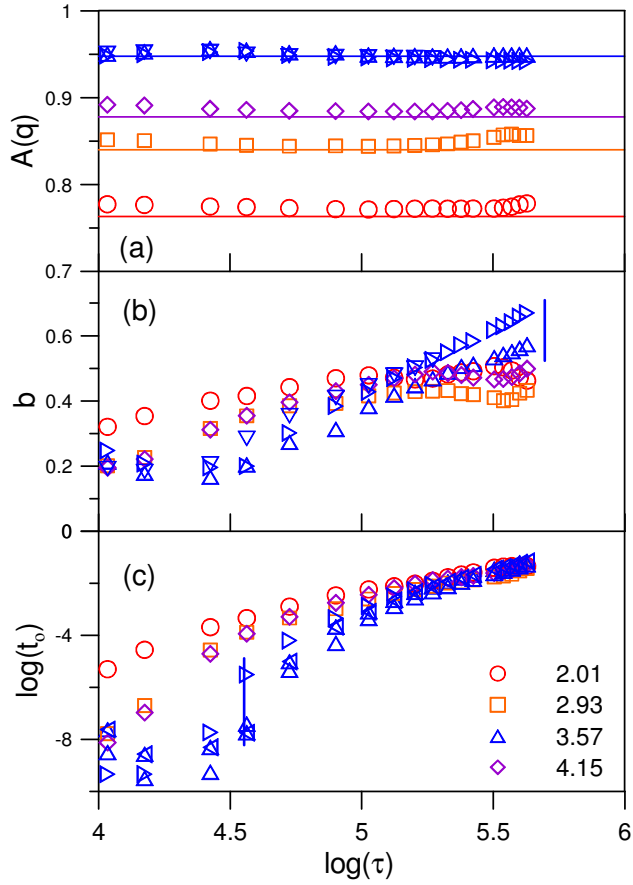

Fig. 11. Parameters, A(q), b and $t_o$ of the power law, Eq. (8), versus logarithm of delay time, for values of qR indicated. Lines are $f(q,\infty)$. For qR=3.57 results of several independent experiments are indicated by triangles in different orientations. Vertical bars are indicative of experimental uncertainty.

Second, variation of the cross-over time, $\tau_m(q)$, with $t_w$ is shown in Fig. 13. So, the stationarity found for the amplitude, $f(q,\tau_m(q))$ above, is not replicated by the other quantities, $\nu(q)$ and $\tau_m(q)$, that characterize the cross-over from fast to slow processes. However, in view of the noise on these data, the possibility that $\tau_m(q)$ ultimately approaches a constant value cannot be excluded. Note also that $\tau_m(q)$ shows no systematic dependence on the wavevector.

We digress to point out that this scale invariance of $\tau_m(q)$ is seen only in the glass and not for the metastable state. Around the freezing value, $\phi \approx \phi_f$, the variation of $\tau_m(q)$ with q is approximately a factor of 10, over the range of q values considered here, and is found to gradually disappear as $\phi$ approaches $\phi_g$ [12].

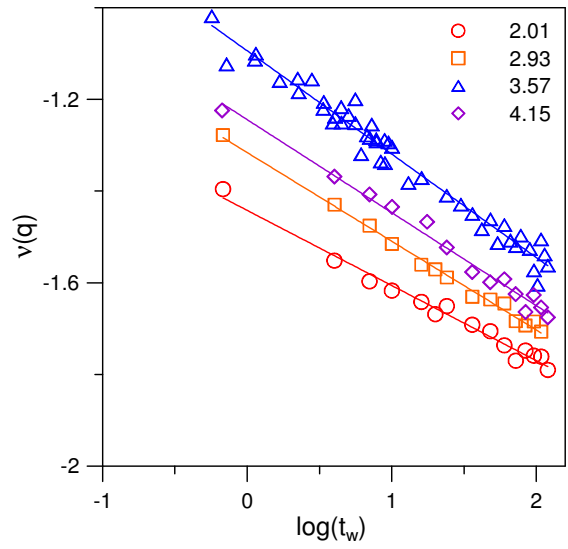

Fig.12. Double logarithm plots of the stretching index, $v_c(q)$, versus waiting time, $t_w$, for values of qR indicated. Lines are power laws, Eq. (9), fitted to the data. The fitting parameters, $\zeta$ and $v_o$, are given in Table 2.

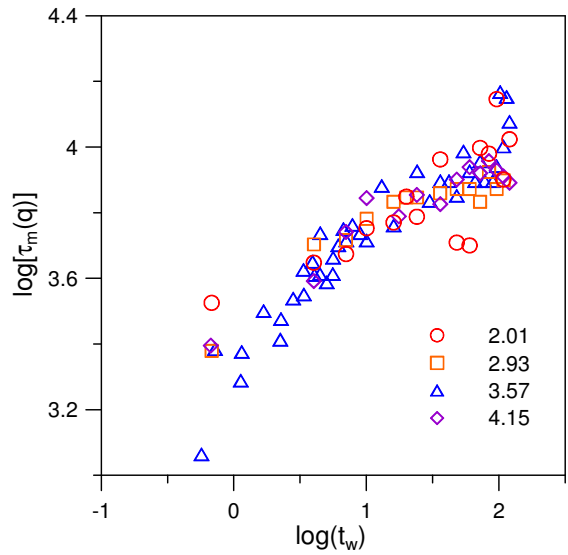

Fig. 13. Double logarithm plot of delay time, $\tau_m(q)$, at the crossover from fast to slow processes, versus waiting time, $t_w$, at indicated values of qR.

| qR | Amplitude $\nu_o$ ± .001 | Ageing exponent $\zeta$ ± 0.02 | $\nu_o/f(q,\infty)$ | $\nu_o/S(q)$ | S(q) PY8% phi=0.6 |
|---|---|---|---|---|---|
| **2.01** | 0.037 | 0.18 | 0.6 | 60 | 0.022 |
| **2.93** | 0.049 | 0.20 | 0.7 | 11 | 0.150 |
| **3.57** | 0.080 | 0.22 | 1.0 | 1.0 | 3.11 |
| **4.15** | 0.056 | 0.20 | 0.7 | 4 | 1.38 |

Table. 2. Values of the fitting parameters for the waiting time dependence of the stretching index (Eq. 9) for XL52 at $\phi$=0.60. The values in the last two columns are normalized to unity at qR=3.57. S(q) has been calculated by the Percus-Yevick approximation [26].

One sees from Table 2, that this apparent lack of sensitivity to the spatial scale, q, also applies to the exponent, $\zeta$, but not to the amplitude, $\nu_o$. So, the latter appears to be the only quantity which characterizes the aging of the crossover from fast to slow processes, that shows any significant dependence on q. The (normalized) ratios $\nu_o/f(q,\infty)$ and $\nu_o/S(q)$ are also listed in Table 2. Despite their appreciable uncertainty, these results indicate that the q-dependence of $\nu_o$ is in much closer harmony with $f(q,\infty)$ than with S(q). So, if there is a correlation of the ageing dynamics with the structure then it is more likely to be with the arrested structure, expressed by $f(q,\infty)$, rather than with the total structure, expressed by S(q).

Finally, we quantify the slow decays of the ISF (fig. 1) in terms of the generalized exponential,

$$f(q,\tau > \tau_m(q)) = A(q)\exp\left[-(\tau/\tau_s)^\delta\right], \qquad (10)$$

We use values for A(q) obtained above, leaving $\tau_s$ and $\delta$ as fitting parameters. These fits, and the fit parameters, are shown in fig. 1. Although this analysis is limited for the present data by the limited decays of the ISFs from their respective plateaux, it is quite clear, from Fig. 1, that $\delta<1$; ie, the time correlation function of the non-stationary processes follow a stretched exponential function of the delay time. So, whatever the nature of the aging dynamics in these colloidal glasses, it differs in some basic respect from ageing characterised by compressed exponential decays with $\delta \approx 1.5$, found in other complex, arrested materials mentioned in Sec. 1. However, due to the limited extent of the decay from the plateau there is insufficient information to quantify the q-dependence of $\tau_s$, nor, for that matter, to preclude the possibility of a crossover to a power-law as observed in computer simulations of molecular glasses [3,4].

## 3.2 Effect of Volume fraction

The results presented in this sub-Section are based on measurements with samples denoted as XL63 in Table 1.

Fig. 14 shows the ISFs obtained by the procedures described in Sec. 2.2 for several volume fractions. The results are qualitatively similar to those in Fig. 1 and 9. Again, the age dependent decays, from the respective plateaux, can be described by stretched exponential function of delay time. As illustrated in Fig. 15 for several volume fractions, the ISF at the crossover, $f(q, \tau_m(q))$, coincides, within experimental noise, with the non-ergodicity factor, $f(q, \infty)$.

Except for the lowest volume fraction, the manner by which the ISFs, $f(q, \tau > \tau_m(q) ; t_w)$, age can be described by the power law, Eq. (8). And as found for the colloidal glass in Sec. 3.1, the amplitude A(q) of the power law agrees with the non-ergodocity factor, $f(q, \infty)$ (see Fig. 16). The decay of the ISFs away from the plateaux, $f(q, \tau_m(q))$, in these cases, occurs entirely by non-stationary processes. The slow decay of the ISF for the sample at $\phi = 0.563$ ($\approx \phi_g$) shows only limited dependence on $t_w$, but once this has exceeded some 10 h, stationarity appears to have been attained. The ISFs for suspensions at lower volume fractions show no dependence on $t_w$ (data not shown). So, as proposed in previous work [25] on these suspensions, there is a narrow range of volume fractions around $\phi_g$ ($\approx 0.563$) where there is a change, from reversible to irreversible, in the processes that effect the final decay of the ISF. This identification of the operational GT is independent of the experimental time window provided, of course, its upper limit exceeds the crossover, $\tau_m(q)$, by a margin sufficient to observe some decay from the plateau. Increasing the time window beyond the present $\tau_{max}$ would coarse grain over the very non-stationary processes we are attempting to resolve.

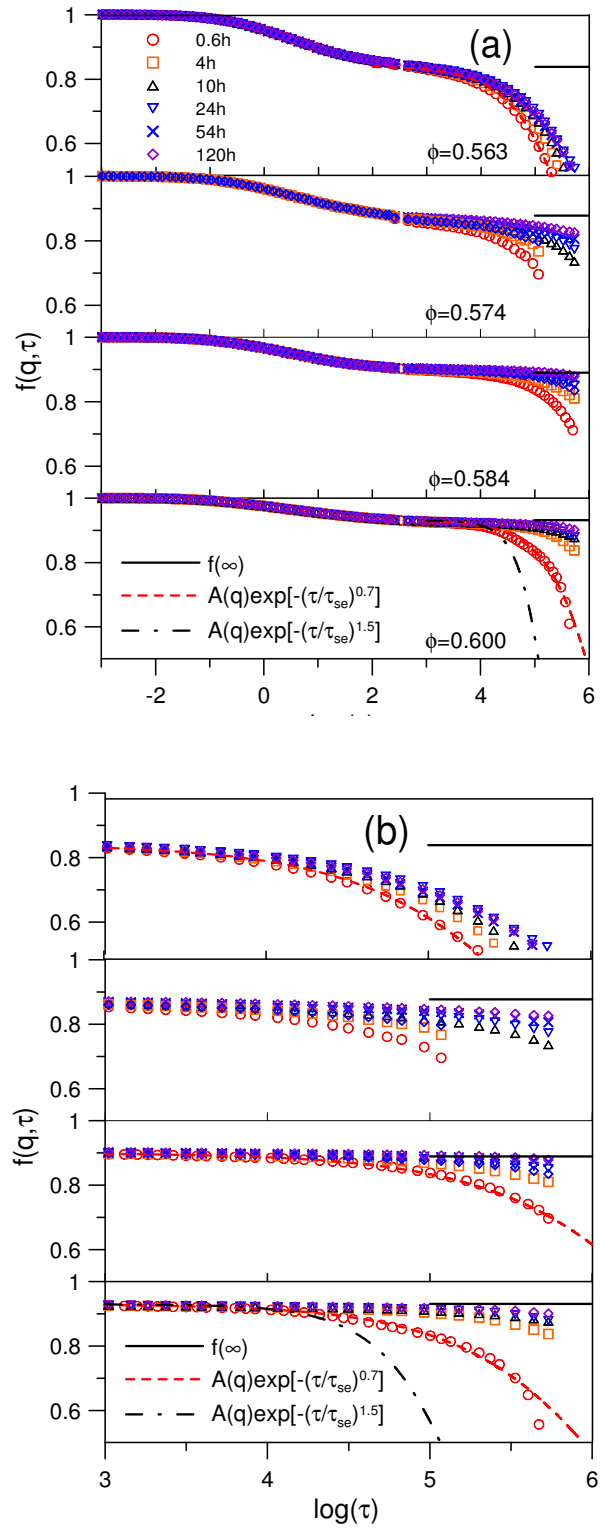

Fig. 14 ISF versus logarithm of delay time, at qR=3.30 for waiting times and volume fractions indicated. (b) Expanded scale of delay time giving clearer fits to stretched exponentials.

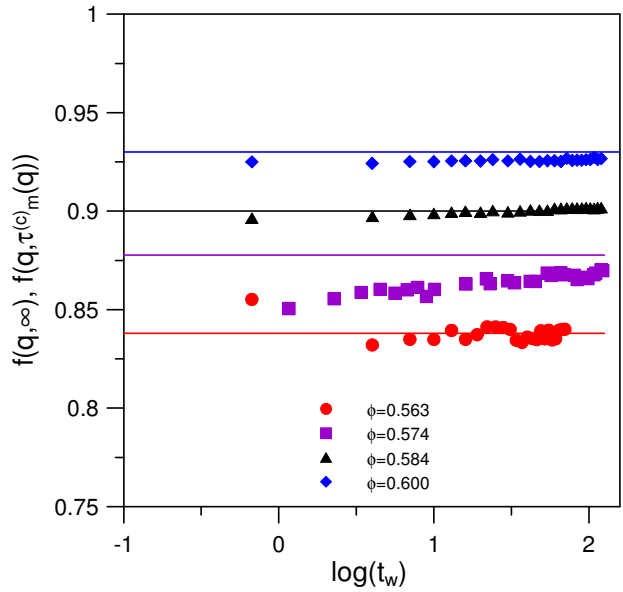

Fig. 15. The ISF, $f(q,\tau_m(q))$, at the cross-over versus logarithm of the waiting time (hours), for volume fractions indicated ($qR=3.30$). The horizontal lines indicate the (average) non-ergodicity factors, $f(q,\infty)$.

The stretching indices, $\nu(q)$, are shown as functions of the waiting time in Fig. 17. For $\phi=0.563$ these show only a weak decrease with waiting time beyond experimental noise. For the remaining cases the waiting time dependence of $\nu(q)$ can be described by the power law given by Eq. (9). As found in Sec. 3.1 for the glass of XL52, the exponents, $\zeta$ tabulated in Table 3 for the highest two volume fractions, show no systematic dependence on wavevector. Moreover, they don't exhibit a detectable dependence on volume fraction. Also as found in Sec. 3.1, the variations of the amplitudes, $\nu_o$, with q are in closer harmony with $f(q,\infty)$ than they are with $S(q)$.

| qR | $\phi=0.584$ | | | $\phi=0.600$ | | |
|---|---|---|---|---|---|---|
| | Amplitude $\nu_o$ $\pm .002$ | Ageing exponent $\zeta$ $\pm 0.02$ | S(q) PY8% | Amplitude $\nu_o$ $\pm .002$ | Ageing exponent $\zeta$ $\pm 0.02$ | S(q) PY8% |
| 1.86 | -- | -- | -- | 0.030 | 0.21 | 0.018 |
| 2.71 | 0.025 | 0.21 | 0.10 | 0.032 | 0.19 | 0.08 |
| 3.30 | 0.040 | 0.21 | 0.90 | 0.056 | 0.23 | 0.70 |
| 3.84 | 0.035 | 0.23 | 2.56 | 0.043 | 0.20 | 2.94 |

Table 3. Values of the fitting parameters for the waiting time dependence of the stretching index (Eq. 9) for two volume fractions of XL63. For $\phi=0.574$, data was only collected at qR=3.30, which yielded $\nu_o$=0.072 and $\zeta$=0.20. S(q) has been calculated by the Percus-Yevick approximation [26].

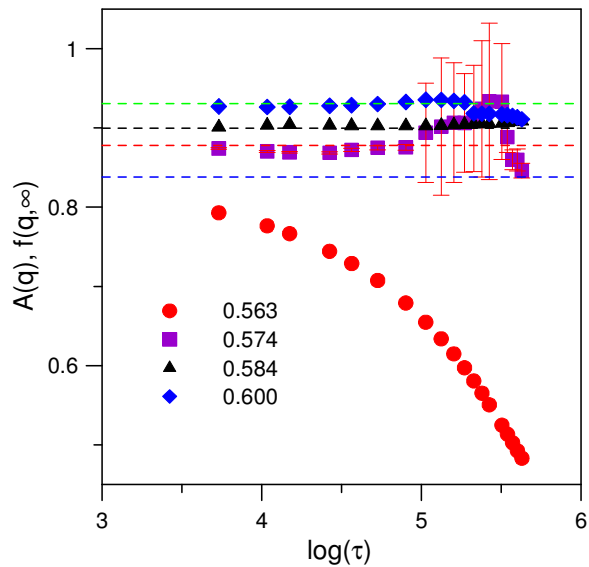

Fig. 16. Parameters, A(q) of the power law, Eq. (8), versus logarithm of delay time, for volume fractions indicated (qR=3.30). Dashed lines are the non-ergodicity factors, $f(q,\infty)$. The large errors for $\phi=0.574$ result from the shorter measurement time $T_2$ of 2000s in this case for smallest waiting times (note also corresponding maximum delay times in the second panel of Fig. 14).

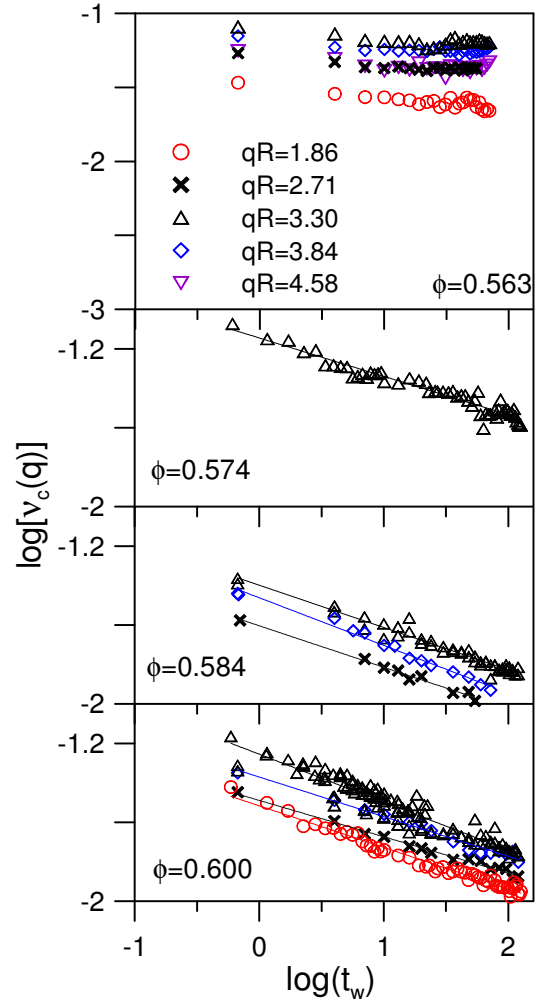

Fig. 17. Double logarithm plots of stretching index, $\nu_c(q)$, versus waiting time, $t_w$, for values of qR and volume fractions indicated. Lines are power laws, Eq. (9), fitted to the data. Values of the fitting parameters, $\zeta$ and $\nu_o$, are given in Table 3.

## 4. CONCLUSIONS

In the above we have reported the results of DLS measurements of the coherent ISF of glasses of colloidal hard spheres for several volume fractions and scattering vectors bracketing the primary peak of the static structure factor. The main results are summarised as follows;

1. In all cases the ISF features a clear crossover, at delay time $\tau_m(q)$, from an initial, predominantly stationary, decay to a slower non-stationary decay.

2. The ISF at the crossover, $f(q,\tau_m(q))$, coincides with the (independently measured) non-ergodicity factor, $f(q,\infty)$.

3. Slow, non-stationary decays, ie. $f(q,\tau>\tau_m(q);t_w)$, insofar as they can be quantified in this work, follow stretched exponential functions of delay time.

4. $f(q,\tau>\tau_m(q)t_w)-f(q,\infty)$ varies with waiting time in a power-law fashion; ie, stationarity is approached algebraically.

In addition to these direct observations of the ageing process itself, the delay time, $\tau_m(q)$, the degree of stretching, $\nu(q)$, and the value, $f(q,\tau_m(q))$, of the ISF at the crossover are perhaps more significant in that their analysis exposes a coupling of ageing to the fastest processes detected in these experiments.

So, while on the face of it, the initial fast decay, $f(q,\tau<\tau_m(q))$, of the ISF exhibits no dependence on the waiting time, $t_w$, ie. it appears to be stationary, it cannot be strictly so because the index, $\nu(q)$, a quantity indicative of the coupling between the fast and the slow, non-stationary processes, is finite. Only for the perfect glass, an idealization approached algebraically with $t_w$, are all processes stationary. Conversely, $\nu(q)$ is a measure of those non-stationary, ergodicity restoring processes absent in the perfect glass. However, the amplitude, $\nu_o$ [Eq. (9)], of the power-law characterizing the ageing of $\nu(q)$, is determined by the final, arrested structure of the ideal glass. So, it comes as no surprise that the manner in which $\nu_o$ varies with wavevector tends to follow the amplitude, $f(q,\infty)$, of the arrested structure rather than that of the total, average structure, $S(q)$.

As far as we can gauge from the range, albeit a limited one, of parameters, $qR$ and $\phi$, covered in these experiments, the index, $\zeta$ of the power law just mentioned shows no significant dependence on either wavevector or volume fraction. Also the crossover time, $\tau_m(q)$, shows no variation with wavevector. The latter should be seen in the context of a previous study [12] of the metastable fluid-like states of these suspensions between the freezing point, $\phi_f$, and the glass transition, $\phi_g$. It was found that the q-dependence of $\tau_m(q)$ varies approximately in harmony with S(q) around $\phi_f$, but that this q-dependence disappears as $\phi_g$ is approached. So as the GT is approached, and density fluctuations become arrested, the irreversible processes that emerge posses a degree of spatial scale invariance. This, along with the algebraic dependence of the index, $\nu(q)$, and the difference, $f(q,\tau>\tau_m(q))-f(q,\infty)$, on waiting time indicate that ageing is an intermittent rather than a continuous process.

Ultimately and however slowly, extraneous effects such as gravity notwithstanding, these colloidal glasses must eventually crystallize. Eventually the ergodicity restoring processes, identified here, must disrupt the arrested structure irreversibly to such an extent that the structure that evolves is able to support a greater spread of propagating modes than does the glass. The facts that these samples remain visibly amorphous and, more specifically, that the fraction of arrested, amorphous structure, $f(q,\infty)$, remains fixed indicates that our observations apply to early stage ageing, ie, before the onset of crystallization.